\documentstyle[aps,prl,twocolumn,floats,epsfig]{revtex}
\newcommand{\Z}{{\sf Z \!\!\! Z}}

\newcommand{\p}{\partial}
\begin{document} \draft 
\title{Study of $CP(N-1)$ $\theta$-Vacua by Cluster-Simulation of $SU(N)$ 
Quantum Spin Ladders}
\author{B.\ B.\ Beard$^\dagger$, M.\ Pepe$^\ddagger$, S.\ Riederer$^\ddagger$,
and U.-J.\ Wiese$^\ddagger$}
\address{$^\dagger$ Departments of Physics and Mechanical Engineering, Christian
Brothers University, Memphis, TN 38104, U.S.A.}
\address{$^\ddagger$ Institute for Theoretical Physics, Bern University,
Sidlerstrasse 5, 3012 Bern, Switzerland}
\date{June 29, 2004}
\twocolumn[\hsize\textwidth\columnwidth\hsize\csname @twocolumnfalse\endcsname
\maketitle

\begin{abstract}
{\em D-theory} provides an alternative lattice regularization of the $(1+1)$-d 
$CP(N-1)$ quantum field theory. In this formulation the continuous classical 
$CP(N-1)$ fields emerge from the {\em dimensional} reduction of {\em discrete}
$SU(N)$ quantum spins. In analogy to Haldane's conjecture, ladders consisting 
of an even number of transversely coupled spin chains lead to a $CP(N-1)$ model
with vacuum angle $\theta = 0$, while an odd number of chains yields $\theta = 
\pi$. In contrast to Wilson's formulation of lattice field theory, in D-theory 
no sign problem arises at $\theta = \pi$, and an efficient cluster algorithm is
used to investigate the $\theta$-vacuum effects. At $\theta = \pi$ there is a 
first order phase transition with spontaneous breaking of charge conjugation 
symmetry for $CP(N-1)$ models with $N>2$.
\end{abstract} 
]

$CP(N-1)$ models are interesting $(1+1)$-d quantum field theories \cite{DAd78} 
which share a number of important features with $(3+1)$-d QCD. These include 
asymptotic freedom, the dynamical generation of a mass-gap, and 
an instanton topological charge leading to non-trivial $\theta$-vacuum effects.
Despite the fact that Nature has chosen $\theta = 0$, it is an interesting 
challenge to understand the structure of $\theta$-vacua, which in QCD may, for 
example, lead to 't Hooft's oblique confinement phases \cite{tHo81}. It has 
been conjectured that $CP(N-1)$ models have a phase transition at $\theta = 
\pi$. In the $CP(1) = O(3)$ case this phase transition is known to be second 
order with a vanishing mass-gap \cite{Hal83,Aff86,Bie95}. For $N>2$, on the 
other hand, the transition is conjectured to be first order \cite{Aff88}, which
is consistent with large $N$ analytic results \cite{Sei84}. For finite $N$, 
however, the investigation of these nonperturbative problems is highly 
nontrivial. In particular, in contrast to $O(N)$ models and other $(1+1)$-d 
quantum field theories, $CP(N-1)$ models with $N>2$ cannot be solved with the 
Bethe ansatz. This is because in the $CP(N-1)$ case an infinite set of 
classical symmetries is anomalous and can thus not be used to solve the quantum
theory analytically.

The numerical investigation of lattice $CP(N-1)$ models is also far from being 
straightforward, even at $\theta = 0$. Again, in contrast to $O(N)$ models 
which can be studied with the efficient Wolff cluster algorithm \cite{Wol89}, 
no efficient cluster algorithm is available for $CP(N-1)$ models \cite{Jan92}. 
There is even a no-go theorem that forbids the construction of an efficient
Wolff-type embedding algorithm for these models \cite{Car93}. Still, at 
$\theta = 0$ a rather efficient multi-grid algorithm was developed in 
\cite{Has92}. However, at $\theta = \pi$ the situation is much worse due to a 
very severe sign problem: the contributions from odd topological charge sectors
almost completely cancel those from even charge sectors. This makes it 
exponentially hard to access large lattices which is necessary for reaching 
reliable conclusions about the phase structure. For this reason, previous 
numerical studies of $\theta$-vacua were limited to moderate volumes 
\cite{Wie89,Bur01} or rely on additional assumptions \cite{Azc02}. In the 
$CP(1) = O(3)$ case, a Wolff-type meron-cluster algorithm allows efficient
simulations at $\theta = \pi$ \cite{Bie95}. Unfortunately, due to the no-go 
theorem this algorithm cannot be extended to higher $CP(N-1)$ models. 

In this paper, we present a method that allows us, for the first time, to 
perform accurate unbiased numerical simulations of any $CP(N-1)$ model at 
$\theta = \pi$. Our method is based on the {\em D-theory} formulation 
of field theory \cite{Cha97} in which continuous classical fields emerge 
dynamically from the {\em dimensional} reduction of {\em discrete} variables. 
D-theory provides an alternative lattice regularization of field theory which 
also yields the same universal continuum theory but is otherwise very different
from Wilson's lattice field theory. In the case of $CP(N-1)$ models, the 
discrete variables are generalized quantum spins $T_x^a = 
\frac{1}{2} \lambda_x^a$ which generate an $SU(N)$ symmetry $[T_x^a,T_y^b] = 
i \delta_{xy} f_{abc} T_x^c$. Here $f_{abc}$ are the structure constants of 
$SU(N)$. The spins are located on the sites $x$ of a square lattice with 
spacing $a$ of size $L \times L'$, with $L \gg L'$ and with periodic boundary 
conditions. Hence, as shown in figure 1, we have a quantum spin ladder 
consisting of $n = L'/a$ transversely coupled spin chains of length $L$. 
\begin{figure}[tb]
\vspace{0.8cm}
\epsfig{file=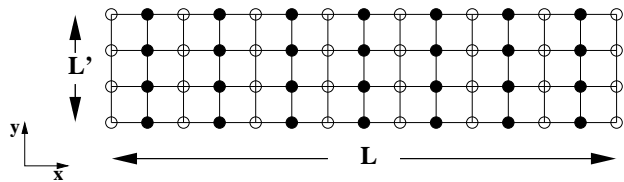,
width=1.2cm,angle=0,bbllx=0,bblly=0,bburx=225,bbury=337}
\vspace{0.2cm}
\caption{\it Spin ladder geometry: the open circles belong to sublattice $A$,
while the filled circles form sublattice $B$.}
\end{figure}
The $x$-direction of size $L$ corresponds to the spatial dimension of the 
target $CP(N-1)$ model, while the extra $y$-dimension of finite extent $L'$ 
will ultimately disappear via dimensional reduction. We consider 
nearest-neighbor couplings which are antiferromagnetic along the chains and 
ferromagnetic between different chains. Hence, the lattice naturally decomposes
into two sublattices $A$ and $B$ with even and odd sites along the 
$x$-direction, respectively. Note that neighboring sites along the transverse 
$y$-direction belong to the same sublattice. The spins $T_x^a$ on sublattice 
$A$ transform in the fundamental representation $\{N\}$ of $SU(N)$, while the 
ones on sublattice $B$ are in the antifundamental representation 
$\{\overline N\}$ and are thus described by the conjugate generators 
$- T_x^{a *}$. The quantum spin ladder Hamiltonian is given by
\begin{eqnarray}
H&=&- J \sum_{x \in A} [T_x^a T_{x+\hat 1}^{a *} + T_x^a T_{x+\hat 2}^a]
\nonumber \\
&-&J \sum_{x \in B} [T_x^{a *} T_{x+\hat 1}^a + T_x^{a *} T_{x+\hat 2}^{a *}],
\end{eqnarray}
where $J > 0$, and $\hat 1$ and $\hat 2$ are unit-vectors in the spatial $x$-
and $y$-directions, respectively. Note that, in the $SU(2)$ case, the $\{2\}$ 
and $\{\overline 2\}$ representations are unitarily equivalent. In particular, 
by a unitary transformation one can turn $- T_x^{a *}$ into $T_x^a$. This
is not possible for $N>2$. By construction the system has a global $SU(N)$ 
symmetry, i.e.\ $[H,T^a] = 0$, with the total spin given by 
\begin{equation}
T^a = \sum_{x \in A} T_x^a - \sum_{x \in B} T_x^{a *},
\end{equation}
which satisfies the $SU(N)$ algebra $[T^a,T^b] = i f_{abc} T^c$.

A priori it is not obvious that the $(2+1)$-d quantum spin ladder provides a 
viable regularization of the $(1+1)$-d continuum $CP(N-1)$ field theory. As a 
necessary prerequisite, the quantum spin ladder does indeed have the global 
$SU(N)$ symmetry of $CP(N-1)$ models. Using the coherent state technique of
\cite{Rea89} one finds that, at zero temperature, the infinite system (with 
both $L, L' \rightarrow \infty$) undergoes spontaneous symmetry breaking from 
$SU(N)$ to $U(N-1)$. It should be noted that the choice of the $SU(N)$ spin 
representations (in this case $\{N\}$ and $\{\overline N\}$) has an influence 
on the breaking pattern. Due to spontaneous symmetry breaking, there are 
massless Goldstone bosons --- in this case spin waves --- described 
by fields in the coset space $SU(N)/U(N-1) = CP(N-1)$. The $CP(N-1)$ fields 
$P(x,y,t)$ are Hermitean $N \times N$ projector matrices, i.e.\ $P^2 = P$, with
$\mbox{Tr} P = 1$. Using chiral perturbation theory, the lowest-order terms in 
the Euclidean effective action for the spin waves are given by
\begin{eqnarray}
\label{action}
S[P]&=&\int_0^\beta dt \int_0^L dx \int_0^{L'} dy \ \mbox{Tr} 
\{\rho_s' \p_y P \p_y P \nonumber \\
&+&\rho_s [\p_x P \p_x P + \frac{1}{c^2} \p_t P \p_t P] -
\frac{1}{a} P \p_x P \p_t P\}.
\end{eqnarray}
Here $\beta = 1/T$ is the inverse temperature, $\rho_s$ and $\rho_s'$ are spin 
stiffness parameters for the $x$- and $y$-direction, respectively, and $c$ is 
the spin wave velocity. The action is invariant under global transformations 
$P' = \Omega P \Omega^\dagger$ with $\Omega \in SU(N)$. The last term in the 
integrand of eq.(\ref{action}) is purely imaginary and is related to the 
topological charge
\begin{equation}
Q[P] = \frac{1}{\pi i} \int_0^\beta dt \int_0^L dx \ 
\mbox{Tr}[P \p_x P \p_t P],
\end{equation}
which is an integer in the second homotopy group $\Pi_2[CP(N-1)] = \Z$, and 
thus $y$-independent. Hence, the $y$-integration in the last term of 
eq.(\ref{action}) can be performed trivially. This yields $i \theta Q[P]$ where
the vacuum angle is given by $\theta = L' \pi/a = n \pi$. Here $a$ is the 
lattice spacing of the quantum spin ladder and $L'/a = n$ is the number of 
transversely coupled spin chains. Hence, for even $n$ the vacuum angle is 
trivial, while for odd $n$ it is equivalent to $\theta = \pi$. The same 
topological term is also generated when one uses a single antiferromagnetic 
spin chain with larger representations of $SU(N)$. When one chooses a 
completely symmetric representation with a Young tableau containing $n$ boxes
in a single row on sublattice $A$ and its anti-representation on sublattice 
$B$, the resulting vacuum angle is again given by $\theta = n \pi$.

While the infinite $(2+1)$-d system has massless Goldstone bosons, the
Coleman-Hohenberg-Mermin-Wagner theorem forbids the existence of massless
excitations once the $y$-direction is compactified to a finite extent $L'$.
As a consequence, the Goldstone bosons then pick up a nonperturbatively
generated mass-gap $m = 1/\xi$ and thus have a finite correlation length $\xi$.
Interestingly, for sufficiently many transversely coupled chains, the 
correlation length becomes exponentially large
\begin{equation}
\xi \propto \exp(4 \pi L' \rho_s/c N) \gg L',
\end{equation}
and the system undergoes dimensional reduction to the $(1+1)$-d $CP(N-1)$
field theory with the action
\begin{eqnarray}
S[P] = \int_0^\beta dt \int_0^L dx \ \mbox{Tr}&\{&\!\!\frac{c}{g^2} 
[\p_x P \p_x P + \frac{1}{c^2} \p_t P \p_t P] \nonumber \\
&-&n P \p_x P \p_t P\}.
\end{eqnarray}
The coupling constant of the dimensionally reduced theory is given by $1/g^2 =
L'\rho_s/c$. The mechanism of dimensional reduction is well-known from quantum
antiferromagnets \cite{Cha88,Has91} and occurs in all D-theory models. It has 
also been discussed for $SU(2)$ quantum spin ladders \cite{Cha96}. The 
dimensional reduction of ladder systems has already been used in the D-theory 
construction of the 2-d $O(3)$ model at non-zero chemical potential. The 
corresponding sign problem has been solved with an efficient meron-cluster 
algorithm \cite{Cha02}. In this paper, we extend the D-theory construction to 
$CP(N-1)$ models, which allows us to simulate them reliably at $\theta = 0$ and
$\pi$. 

One advantage of D-theory is that it allows us to construct quantum field 
theories using simple discrete degrees of freedom instead of the usual
continuum classical fields. In particular, the partition function 
$Z = \mbox{Tr} \exp(- \beta H)$ of the $SU(N)$ quantum spin ladder can be
written as a path integral using a basis of discrete $SU(N)$ spin states 
$q \in \{u,d,s,...\}$ on sublattice $A$ and $\overline q \in \{\overline u,
\overline d, \overline s,...\}$ on sublattice $B$. For $SU(2)$ this 
corresponds to the usual $\uparrow$ and $\downarrow$ spins. These can be 
simulated with the very efficient loop-cluster algorithm \cite{Eve93,Wie94} 
which can even operate directly in continuous Euclidean time \cite{Bea96}. As
mentioned in \cite{Kaw03}, this algorithm extends to $SU(N)$ in a 
straightforward manner.

We have used this cluster algorithm to investigate if $CP(N-1)$ models with
$N>2$ have a first order phase transition at $\theta = \pi$ where the charge 
conjugation symmetry $C$ is spontaneously broken. A natural quantity that 
suggests itself as an order parameter for such a phase transition would be the 
topological charge $Q[P]$ which indeed is $C$-odd, i.e.\ $Q[^CP] = Q[P^*] = 
- Q[P]$. Note that $C$ is explicitly broken for $\theta \neq 0,\pi$. At 
$\theta = \pi$ it is not broken explicitly because then the
Boltzmann weight $\exp(i \theta Q[P]) = (-1)^{Q[P]}$ is $C$-invariant. The
topological charge itself is defined only in the framework of the target
continuum theory. In the discrete spin system another order parameter, which is
also $C$-odd, is more easily accessible. In the basis of quantum spin states 
$q,\overline q$ we define an order parameter $Q[q,\overline q]$ by counting
the number of spin flips in a configuration. $Q[q,\overline q]$ receives a 
contribution 1 if a pair of nearest neighbor spins along the $x$-direction, 
$q_x \overline q_{x+\hat 1}$, flips to another state, 
$q_x' \overline q_{x+\hat 1}'$, at some moment in time. A spin flip from 
$\overline q_x q_{x+\hat 1}$ to $\overline q_x' q_{x+\hat 1}'$, on the other 
hand, contributes $-1$ to $Q[q,\overline q]$. In the quantum spin ladder, 
charge conjugation corresponds to replacing each spin state $q_x$ by 
$q_{x+\hat 1}$ (which is the conjugate of $\overline q_{x+\hat 1}$). Indeed, 
$Q[q,\overline q]$ changes sign under this operation while the action remains 
invariant.

We have used the cluster algorithm to simulate $SU(N)$ quantum spin ladders for
$N = 3,4$, and 5, with $n = L'/a \in \{2,3,...,7\}$. The spatial size $L/a$ has
been varied between 20 and 300, and the inverse temperature $\beta J$ was 
chosen between 15 and 200. Our simulations confirm the existence of a first 
order phase transition with spontaneous $C$-breaking at $\theta = \pi$ for all 
$N>2$. As expected, there is no phase transition at $\theta = 0$. Figure 2 
shows Monte Carlo time histories of $Q[q,\overline q]$ for $SU(4)$ spin ladders
with $n = 3$ and 4 which correspond to $CP(3)$ models at $\theta = \pi$ and 0, 
respectively. For $n = 3$ one observes two coexisting phases with spontaneous 
$C$-breaking, while for $n = 4$ there is only one phase which is $C$-symmetric.
For $n = 5$ and thus $\theta = \pi$ one again finds a first order phase 
transition as in the $n = 3$ case.
\begin{figure}[tb]
\epsfig{file=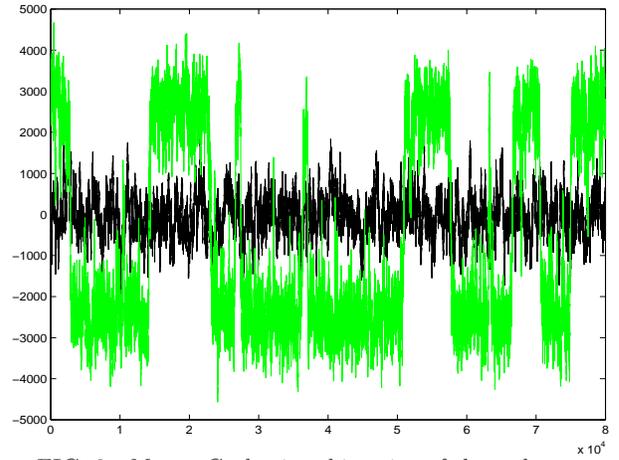,width=8.cm,height=6.1cm,angle=0}
\caption{\it Monte Carlo time histories of the order parameter
$Q[q,\overline q]$ for the $CP(3)$ model at $\theta = 0$ 
($n=4$, $L=180 a$, $\beta J = 50$) and $\theta = \pi$ 
($n=3$, $L=200 a$, $\beta J = 80$).}
\end{figure}
Figure 3 shows the probability distribution of $Q[q,\overline q]$ for an 
$SU(5)$ spin ladder with $n = L'/a = 7$, which corresponds to a $CP(4)$ model 
at $\theta = \pi$. The double peak structure again implies a first order phase
transition. A first order phase transition is also observed in the $CP(2)$
case.
\begin{figure}[tb]
\vspace{-.4cm}
\hspace{1.3cm}
\epsfig{file=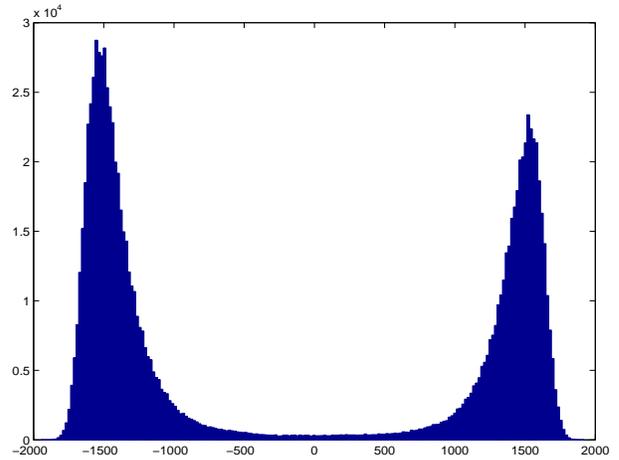,width=8.cm,height=6.1cm,angle=0}
\vspace{.1cm}
\caption{\it Probability distribution of the order parameter
$Q[q,\overline q]$ for the $CP(4)$ model at $\theta = \pi$ 
($n\! =\!7$, $L\! =\! 40 a$, $\beta J\! =\! 6$). Due to limited 
statistics the two peaks are sampled unevenly.}
\vspace{-.3cm}
\end{figure}

At this point, we have shown that D-theory indeed verifies the conjectured 
$\theta$-vacuum structure of $CP(N-1)$ models. In order to demonstrate that we
approach the continuum limit of an asymptotically free field theory we have 
determined the correlation length $\xi(n)$ (defined using the second moment 
method~\cite{Car93}) in the $CP(2)$ case as a function of
$n = L'/a$ which controls the coupling $1/g^2 = L' \rho_s/c$. One obtains 
$\xi(2) = 4.82(4) a$, $\xi(4) = 17.6(2) a$, and $\xi(6) = 61(2) a$, which indeed 
shows the exponential increase of the correlation length characteristic for an 
asymptotically free theory.

In D-theory $CP(N-1)$ models at $\theta = 0$ can also be obtained by 
dimensional reduction of an $SU(N)$ quantum ferromagnet with the Hamiltonian
\begin{equation}
H = - J \sum_{x,\mu} T_x^a T_{x+\hat\mu}^a.
\end{equation}
In this case, using a symmetric $SU(N)$ representation with a Young tableau
containing $n$ boxes, the symmetry again breaks spontaneously to $U(N-1)$ and 
the low-energy effective action for the spin waves takes the form
\begin{eqnarray}
S[P]&=&\int_0^\beta \!\!\! dt \int \!\! d^2x \ 
\mbox{Tr}[\rho_s \p_\mu P \p_\mu P -
\frac{2 n}{a^2} \int_0^1 \!\!\! d\tau \ P \p_t P \p_\tau P] \nonumber \\
&\rightarrow&\beta \rho_s \int d^2x \ \mbox{Tr}[\p_\mu P \p_\mu P],
\end{eqnarray}
where $x$ now represents a 2-d space-time coordinate.
The Wess-Zumino term with the quantized prefactor $n$ involves an interpolated
field $P(x,t,\tau)$ ($\tau \in [0,1]$) with $P(x,t,1) = P(x,t)$ and $P(x,t,0) =
\mbox{diag}(1,0,...,0)$. The Wess-Zumino term vanishes after dimensional 
reduction and one obtains the action of the 2-d $CP(N-1)$ model at 
$\theta = 0$ with $1/g^2 = \beta \rho_s$. In order to verify explicitly that 
this defines the $CP(N-1)$ model in the continuum limit, we now compare 
physical results obtained with the $SU(N)$ quantum ferromagnet and the Wilson 
formulation using the standard lattice action
\begin{equation}
S[P] = - \frac{2}{g^2} \sum_{x,\mu} \mbox{Tr} [P_x P_{x+\hat\mu}].
\end{equation}
The continuum limit is reached as $g \rightarrow 0$. A convenient 
physical quantity characteristic for a given model is the universal finite-size
scaling function $F(z) = \xi(2 L)/\xi(L)$. Here $\xi(L)$ is the correlation 
length in a finite system of size $L$ (again obtained from the second moment 
method), and $z = \xi(L)/L$ is a finite size scaling variable that measures the
size of the system in physical units.
\begin{figure}[bt]
\hspace{-.2cm}
\epsfig{file=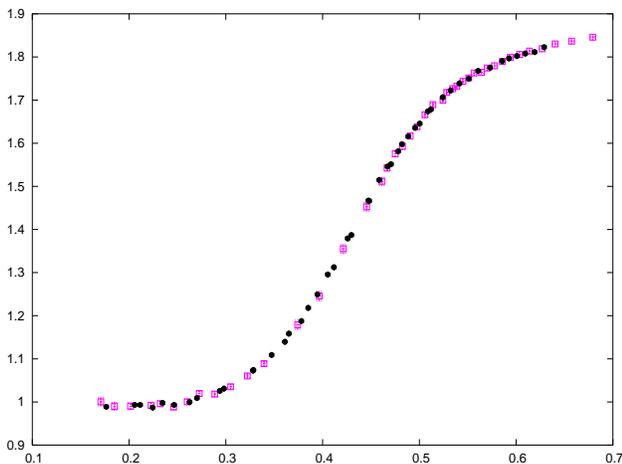,height=8.7cm,width=6.3cm,angle=-90}
\vspace{-0.1cm}
\caption{\it Monte Carlo data for the universal finite-size scaling function 
$F(z)$ of the 2-d $CP(2)$ model at $\theta = 0$. The black points represent 
D-theory data from an $SU(3)$ quantum ferromagnet at $\beta J = 5.5$ and $6$, while the 
open squares correspond to data obtained with the standard Wilson lattice field 
theory at $1/g^2 = 2.25$ and $2.5$.}
\vspace{-0.4cm}
\end{figure}
Figure 4 shows Monte Carlo data for $F(z)$ obtained both from D-theory and from
the standard Wilson approach. Up to small scaling violations, the agreement of 
the two data sets confirms that the $SU(N)$ quantum ferromagnet indeed 
provides a valid lattice regularization of 2-d $CP(N-1)$ models. Thanks 
to the cluster algorithm, the D-theory framework allows calculations that are 
much more accurate than the ones using Wilson's approach.

To summarize, D-theory provides us with a powerful algorithmic tool, an
efficient cluster algorithm that even works at $\theta = \pi$. Despite great 
efforts, this remains impossible within Wilson's framework. In fact, a no-go 
theorem forbids the construction of an efficient Wolff-type embedding 
algorithm. In the D-theory framework this theorem is evaded due to the use of
discrete quantum variables instead of continuous classical fields. Also in 
Wilson's $SU(N)$ lattice gauge theory all attempts to construct efficient 
cluster algorithms have failed. The present study of $(1+1)$-d $CP(N-1)$ models
raises hopes that the D-theory formulation of $(3+1)$-d QCD \cite{Cha97} could 
also be investigated with efficient cluster algorithms. In this way, even 
exotic phenomena like oblique confinement \cite{tHo81} may eventually become 
accessible to numerical investigation.

U.-J.\ W. likes to thank K.\ Holland and B.\ Scarlet for numerous discussions 
on $CP(N-1)$ models in D-theory. We have also benefitted from discussions with
G.\ Colangelo, P.\ Hasenfratz, and F.\ Niedermayer. This work was supported in part by the 
Schweizerischer Nationalfonds.

\end{document}